\documentclass[prx,aps,twocolumn,showpacs,floatfix]{revtex4-1}
\usepackage{tabularx}
\usepackage{amsmath}
\usepackage{amssymb}
\usepackage{hyperref}
\usepackage{graphicx}
\usepackage{dcolumn}
\usepackage{ulem}
\usepackage{bm}

\begin{document}

\title{FFLO or Majorana superfluids: The fate of fermionic cold atoms in
spin-orbit coupled optical lattices}
\author{Chunlei Qu$^{1}$}
\author{Ming Gong$^{1}$}
\author{Chuanwei Zhang$^{1}$}
\thanks{Corresponding author, Email:chuanwei.zhang@utdallas.edu}

\begin{abstract}
The recent experimental realization of spin-orbit coupling (SOC) for
ultra-cold atoms opens a completely new avenue for exploring new quantum
matter. In experiments, the SOC is implemented simultaneously with a Zeeman
field. Such spin-orbit coupled Fermi gases are predicted to support Majorana
fermions with non-Abelian exchange statistics in one dimension (1D).
However, as shown in recent theory and experiments for 1D spin-imbalanced
Fermi gases, the Zeeman field can lead to the long-sought
Fulde-Ferrell-Larkin-Ovchinnikov (FFLO) superfluids with non-zero momentum
Cooper pairings, in contrast to the zero momentum pairing in Majorana
superfluids. Therefore a natural question to ask is which phase, FFLO or
Majorana superfluids, will survive in spin-orbit coupled Fermi gases in the
presence of a large Zeeman field. In this paper, we address this question by
studying the mean field quantum phases of 1D (quasi-1D) spin-orbit coupled
fermionic cold atom optical lattices.
\end{abstract}
\affiliation{$^{1}$ Department of Physics, The University of Texas at Dallas, Richardson,
TX, 75080 USA}
\maketitle
\thanks{Email: chuanwei.zhang@utdallas.edu}

\section{Introduction}

\label{sec: intro} Spin-orbit (SO) coupling plays an important role in many
important condensed matter phenomena \cite{Review1, Review2}, ranging from
spintronics to topological insulators. The recent experimental breakthrough
in realizing SO coupling in ultra-cold Bose and Fermi gases \cite%
{Ian,Pan,WSU,JZhang,Zwierlein} provides a new platform for engineering many
new many-body quantum matters \cite{SOreview}. In the experiments, the SO
coupling is realized together with a Zeeman field. It is well known that
such SO coupling and Zeeman field, together with the \textit{s}-wave
superfluid pairing in degenerate Fermi gases, can support zero-energy
Majorana fermions \cite{Zhang} with non-Abelian exchange statistics when the
Zeeman field is beyond a certain critical value \cite{Kitaev, Nayak}. In
solid state, the same ingredient has been realized using a heterostructure
composed of a semiconductor nanowire (or thin film) with strong SO coupling,
an \textit{s}-wave superconductor, and a magnetic field (or a magnetic
insulator) \cite{Fu, Jay, Alicea, Lutchyn, Oreg, Jay2, Mao, Lee}. Important
experimental progresses have been made along this direction \cite{MFs1,
MFs2, MFs3, MFs4}, where some signatures which may be related with Majorana
fermions have been observed. In the solid state heterostructure, the \textit{%
s}-wave pairing is induced to the semiconductor through proximity effects
\cite{Prox1, Prox2}.

In degenerate fermi gases, however, as observed in both theory and
experiments \cite{Torma, Loh, FFLOexp}, the presence of a large Zeeman field
(realized by the spin population imbalance) can induce non-zero momentum
Cooper pairings between atoms, i.e., FFLO phases \cite{FF64,LO64,LO65},
especially in low-dimensional Fermi gases. Such FFLO phases may not support
Majorana fermions. Therefore it is natural to ask whether the FFLO
superfluids \cite{Iskin1,ZZ, HH, Dong, XJLiu, XJLiu2, WeiYi, Xu} with
non-zero momentum Cooper pairs or the Majorana superfluids with zero
momentum Cooper pairs will survive in the presence of SO coupling and a
large Zeeman field. This question becomes especially important because of
the recent experimental realization of SO coupled Fermi gases \cite{JZhang,
Zwierlein}, which makes the observation of Majorana fermions in cold atomic
systems tantalizingly close. The cold atom system may be a better platform
for the observation of Majorana fermions because of the lack of disorder and
impurity \cite{Jiang, XJLiu, TSclass2, Gong2,Sato, Zhu, Gong}, an issue that
has led to intensive debate in the condensed matter community on the
zero-bias peak signature of Majorana fermions in recent transport
experiments \cite{MFs1,MFs2}.

In this paper, we address the competition between FFLO and Majorana
superfluids by studying the quantum phases of spin-imbalanced Fermi gases in
spin-orbit coupled optical lattices. Because of the fact that the
experimentally realized SO coupling is one-dimensional (1D) and the natural
dimension requirement for the realization of Majorana fermions in this
system, we consider 1D SO coupled optical lattices (similar to 1D nanowires)
and investigate the quantum phase diagram at zero temperature using the mean
field theory. The quantum phases of Fermi atoms are obtained by
self-consistently solving the corresponding Bogoliubov-de Gennes (BdG)
equation. Without SO coupling, there are no Majorana superfluids, and FFLO
superfluids appear in the large Zeeman field region. The SO coupling
enhances the Majorana superfluid phase while suppresses the FFLO superfluid
phase. Majorana and FFLO superfluids exist for different filling factors
(i.e., in different chemical potential region). We characterize different
quantum phases by visualizing their real space superfluid order parameters,
density distributions, and Majorana zero energy wavefunctions. The effects
of the harmonic trap are also discussed. We find that the same quantum
phases are preserved in a 3D optical lattice with weak tunnelings along two
transversal directions (quasi-1D geometry).

The rest of the paper is organized as follows. In Sec. \ref{sec:model}, we
present the BdG equation for describing Fermi atoms in spin-orbit coupled
optical lattices. The symmetry of the BdG equation and its consequence to
the phase diagram is discussed. In Sec. \ref{sec:result}, we present the
main numerical results obtained by self-consistently solving the BdG
equation. We discuss the phase diagram, the characterization of various
phases, and the effects of the harmonic trap and Hartree shift. We also
present the results in a 3D optical lattice with weak tunnelings along two
transversal directions. Sec. \ref{sec: summary} consists of discussion and
conclusion.

\section{Hamiltonian and symmetry}

\label{sec:model} We consider a 1D degenerate Fermi gas with Zeeman field
and SO coupling in an optical lattice. The dynamics of this system can be
described by the standard tight-binding Hamiltonian%
\begin{equation}
\mathcal{H}=\mathcal{H}_{0}+\mathcal{H}_{Z}+\mathcal{H}_{SO},  \label{H-0}
\end{equation}%
where the first term is the usual spin-1/2 Fermi-Hubbard model in an optical
lattice,
\begin{equation}
\mathcal{H}_{0}=-t\sum_{i\sigma }(\hat{c}_{i\sigma }^{\dagger }\hat{c}%
_{i+1\sigma }+H.c.)-\mu \sum_{i\sigma }\hat{n}_{i\sigma }-U\sum_{i}\hat{n}%
_{i\uparrow }\hat{n}_{i\downarrow },
\end{equation}%
$t$ is the hopping amplitude, $\mu $ is the chemical potential, and $U$ is
the contact interaction. The second and third terms are the Zeeman field
\begin{equation*}
\mathcal{H}_{Z}=-h\sum_{i}(\hat{c}_{i\uparrow }^{\dagger }\hat{c}_{i\uparrow
}-\hat{c}_{i\downarrow }^{\dagger }\hat{c}_{i\downarrow })
\end{equation*}%
and SO coupling
\begin{equation*}
\mathcal{H}_{SO}=\alpha \sum_{i}(\hat{c}_{i-1,\uparrow }^{\dagger }\hat{c}%
_{i\downarrow }-\hat{c}_{i+1,\uparrow }^{\dagger }\hat{c}_{i\downarrow
}+H.c.).
\end{equation*}%
Such type of SO coupling and Zeeman field have been realized in recent
experiments for both bosons and fermions \cite{Ian, Pan, WSU, JZhang,
Zwierlein}. In experiments, the hopping amplitude, Zeeman field, SO coupling
strength, and contact interactions may be tuned independently. In our
numerical simulation we set $t=1$ throughout this work. All other energies
are scaled by $t$. The total length of the 1D optical lattice is chosen as $%
N=100$, which is long enough to ensure that the coupling between two ends is
vanishingly small. An open boundary condition is used to obtain the
zero-energy Majorana fermions at two ends of the 1D optical lattice in the
topological superfluid regime.

As the first approach for understanding the quantum phases of such 1D SO
coupled optical lattices, we consider the standard mean field theory. We
decouple the interaction term in $H_{0}$ using
\begin{eqnarray*}
-U\hat{n}_{i\uparrow }\hat{n}_{i\downarrow } &=&\Delta _{i}\hat{c}%
_{i\uparrow }^{\dagger }\hat{c}_{i\downarrow }^{\dagger }+\Delta _{i}^{\ast }%
\hat{c}_{i\downarrow }\hat{c}_{i\uparrow }-|\Delta _{i}|^{2}/U \\
&&+U\langle \hat{n}_{i\uparrow }\rangle \hat{n}_{i\downarrow }+U\hat{n}%
_{i\uparrow }\langle \hat{n}_{i\downarrow }\rangle -U\langle \hat{n}%
_{i\uparrow }\rangle \langle \hat{n}_{i\downarrow }\rangle .
\end{eqnarray*}%
Notice that here we have taken into account the Hartree shift term, which
has quantitative effects on our results. The effective Hamiltonian reads as
\begin{eqnarray}
\mathcal{H}^{\text{eff}} &=&-t\sum_{i}\sum_{\sigma }(\hat{c}_{i\sigma
}^{\dagger }\hat{c}_{i+1\sigma }+H.c.)-\sum_{i\sigma }\tilde{\mu}_{i\sigma }%
\hat{c}_{i\sigma }^{\dagger }\hat{c}_{i\sigma }  \notag \\
&+&\sum_{i}(\Delta _{i}\hat{c}_{i\uparrow }^{\dagger }\hat{c}_{i\downarrow
}^{\dagger }+\Delta _{i}^{\ast }\hat{c}_{i\downarrow }\hat{c}_{i\uparrow })+%
\mathcal{H}_{Z}+\mathcal{H}_{SO},  \label{H-eff}
\end{eqnarray}%
where the chemical potential $\tilde{\mu}_{i\sigma }=\mu +U\langle \hat{n}_{i%
\bar{\sigma}}\rangle $ becomes site dependent, $\bar{\sigma}=-\sigma $. The
superfluid pair potential is defined as $\Delta _{i}=-U\langle \hat{c}%
_{i\downarrow }\hat{c}_{i\uparrow }\rangle $. Using the Bogoliubov
transformation $\hat{c}_{i\sigma }=\sum_{n}(u_{i\sigma }^{n}\hat{\Gamma}
_{n}-\sigma v_{i\sigma }^{n}\hat{\Gamma} _{n}^{\dagger })$, we obtain the
BdG equation
\begin{equation}
\sum_{j}%
\begin{pmatrix}
H_{ij\uparrow } & \alpha _{ij} & 0 & \Delta _{ij} \\
-\alpha _{ij} & H_{ij\downarrow } & -\Delta _{ij} & 0 \\
0 & -\Delta _{ij}^{\ast } & -H_{ij\uparrow } & -\alpha _{ij} \\
\Delta _{ij}^{\ast } & 0 & \alpha _{ij} & -H_{ij\downarrow }%
\end{pmatrix}%
\begin{pmatrix}
u_{j\uparrow }^{n} \\
u_{j\downarrow }^{n} \\
-v_{j\uparrow }^{n} \\
v_{j\downarrow }^{n}%
\end{pmatrix}%
=E_{n}%
\begin{pmatrix}
u_{j\uparrow }^{n} \\
u_{j\downarrow }^{n} \\
-v_{j\uparrow }^{n} \\
v_{j\downarrow }^{n}%
\end{pmatrix}%
,  \label{BdG}
\end{equation}%
where $H_{ij\uparrow }=-t\delta _{i\pm 1,j}-(\tilde{\mu}_{i\sigma }+h)\delta
_{ij}$, $H_{ij\downarrow }=-t\delta _{i\pm 1,j}-(\tilde{\mu}_{i\sigma
}-h)\delta _{ij}$, $\alpha _{ij}=(j-i)\alpha \delta _{i\pm 1,j}$,
\begin{equation}
\Delta _{ij}=-U\delta _{ij}\sum_{n=1}^{2N}[u_{i\uparrow }^{n}v_{i\downarrow
}^{n\ast }f(E_{n})-u_{i\downarrow }^{n}v_{i\uparrow }^{n\ast }f(-E_{n})],
\label{pairing}
\end{equation}%
\begin{eqnarray}
\left\langle \hat{n}_{i\uparrow }\right\rangle
&=&\sum_{n=1}^{2N}[|u_{i\uparrow }|^{2}f(E_{n})+|v_{i\uparrow
}|^{2}f(-E_{n})],  \notag \\
\left\langle \hat{n}_{i\uparrow }\right\rangle
&=&\sum_{n=1}^{2N}[|u_{i\downarrow }|^{2}f(E_{n})+|v_{i\downarrow
}|^{2}f(-E_{n})],  \label{particle}
\end{eqnarray}%
with the Fermi-Dirac distribution $f(E)=1/\left( 1+e^{E/T}\right) $. The BdG
equation (\ref{BdG}) should be solved self-consistently with the order
parameter equation (\ref{pairing}) and the particle number equation (\ref%
{particle}) for the average number of atoms per lattice site $%
n=\sum_{i,\sigma }\left\langle \hat{n}_{i\sigma }\right\rangle /N$. Here we
denote $n$ as the filling factor for convenience (note that generally the
filling factor is defined as $\nu =n/2$ in the literature). In our
simulation, we take the attractive interaction strength $U=4.5$ and the
temperature $T=0$.

The original Hamiltonian Eq. (\ref{H-0}) contains no imaginary part,
therefore the wavefunction in Eq. (\ref{BdG}) can be made real and the order
parameter in Eq. (\ref{H-eff}) is also real. The topological symmetry of the
effective model belongs to BDI class, with characteristic index $\mathcal{Z}$%
, instead of the BdG D-class with $\mathcal{Z}_{2}$ index \cite{TSclass1,
TSclass2}. The mean field Hamiltonian (\ref{H-eff}) still preserves the
basic symmetry properties of Eq. (\ref{H-0}). Consider the particle-hole
operation $\mathcal{C}%
\begin{pmatrix}
\hat{c}_{i\uparrow } \\
\hat{c}_{i\downarrow }%
\end{pmatrix}%
\mathcal{C}^{-1}=(-1)^i
\begin{pmatrix}
\hat{c}_{i\uparrow }^{\dagger } \\
\hat{c}_{i\downarrow }^{\dagger }%
\end{pmatrix}%
$, we have $\mathcal{C}H(\mu )\mathcal{C}^{-1}=H(-\mu )$ when the Hartree
shift term is ignored and $\mathcal{C}H(\mu )\mathcal{C}^{-1}=H(-U-\mu )$
when the Hartree shift term is included. Thus the spectrum should be
symmetric about $\mu =0$ and $\mu=-\frac{U}{2}$ respectively, which is
confirmed in our numerical results (see Fig. 1 and Fig. 6). However, in the
presence of a trapping potential, the chemical potential becomes site
dependent, thus the particle-hole symmetry $\mathcal{C}$ is broken and the
band structure is no longer symmetric about $\mu =0$ (see Fig. 5).

\begin{figure}[tbp]
\centering
\includegraphics[width=3.2in]{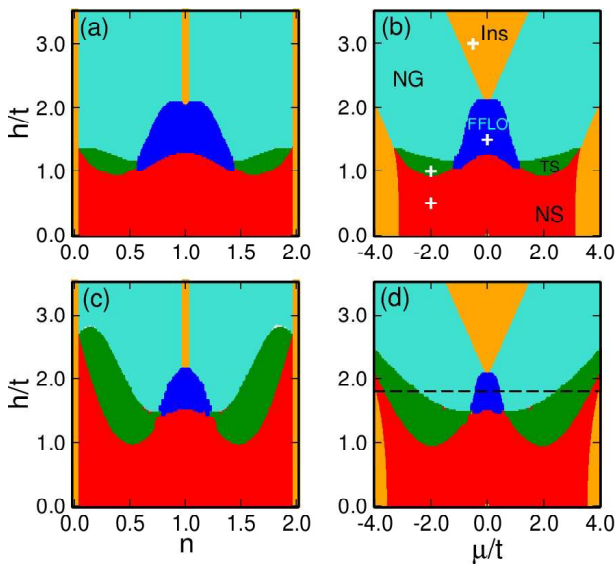}
\caption{(Color online) Phase diagram of 1D SO coupled optical lattices as a
function of Zeeman field $h$ and filling factor $n$ (a,c) or chemical
potential $\protect\mu $ (b,d). First row: the spin orbit coupling $\protect%
\alpha =0.5$, second row: $\protect\alpha =1.0$. Five different phases are
identified in each phase diagram: normal BCS superfluid (NS), topological
superfluid (TS), FFLO, normal gas (NG), and insulator phase (Ins).}
\label{fig-noHS}
\end{figure}

\section{Phase diagram: Majorana versus FFLO superfluids}

\label{sec:result} We self-consistently solve the BdG equations (\ref{BdG},%
\ref{pairing},\ref{particle}) with an open boundary condition to obtain the
phase diagram. The SO coupled optical lattice supports several different
phases: normal BCS superfluid (NS) with $\Delta \neq 0$ and all non-zero
eigenstates; topological superfluids with $\Delta \neq 0$ and zero-energy
Majorana fermions located at two ends of the lattice; FFLO phase with
oscillating $\Delta $ and magnetization; insulator phase (Ins) with integer
filling factor and finite energy gaps; and normal gas (NG) phase without
pairing and energy gap. We first study the phases without Hartree shift, and
address the role of Hartree shift at the end of the section.

\subsection{Phase diagram without Hartree shift}

Our numerical results are presented in Fig. \ref{fig-noHS} for two different
sets of SO coupling strength. In Figs. \ref{fig-noHS} (a) and (c) we plot
the phase diagram in the $h-n$ plane and in Figs. \ref{fig-noHS} (b) and (d)
we plot the results in the $h-\mu $ plane. We see the phase diagram is
symmetric around $n=1$ or $\mu =0$, as discussed in the previous section.
When the Zeeman field $h$ is very small, the system favors the normal BCS
superfluids. With increasing Zeeman field $h$, topological superfluids and
FFLO phases emerge as the ground states of the system for different filling
factors. The FFLO phase is more likely to be observed around the integer
filling factor $n=1$. When the Zeeman field becomes even larger, where only
atoms with one type of spin can stay in each lattice site, the insulator
phase develops with the filling factor $n=1$. The insulator phase can also
be found when $\mu $ is too large (fully occupied band) or to small (empty
band). The topological phase and associated Majorana fermions emerge with
fractional filling factors $n$. By comparing the phase region for different
SO coupling, we find that the SO coupling enhances the topological
superfluid phase and suppresses the FFLO phase. Note that without SO
coupling, there is only FFLO phase, and no topological superfluid phase \cite%
{Torma, Loh}.

\begin{figure}[tbp]
\centering
\includegraphics[width=3.2in]{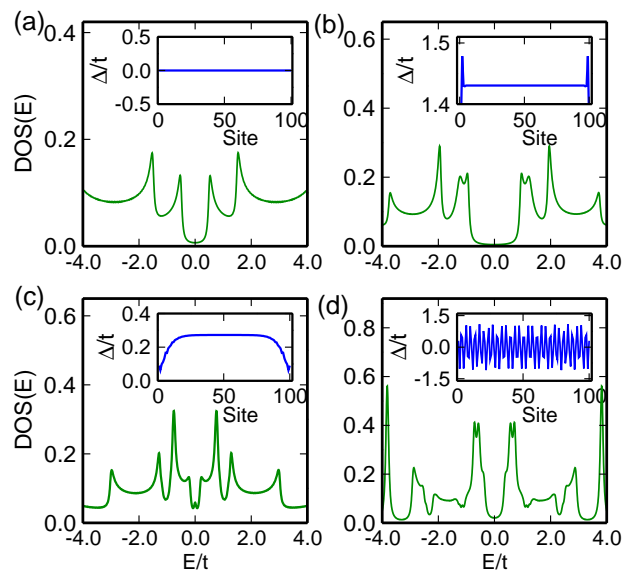}
\caption{(Color online) Representative density of states for (a) insulator
phase, (b) normal BCS superfluid phase, (c) topological superfluid phase,
and (d) FFLO phase. The insets show the order parameters in each phase. The
corresponding phase points are marked by the plus signs in Fig. \protect\ref%
{fig-noHS}b. Note that there is a zero-energy peak in the DOS in the
topological superfluid phase as shown in (c). }
\label{fig-dos}
\end{figure}
\begin{figure}[tbp]
\centering
\includegraphics[width=3.2in]{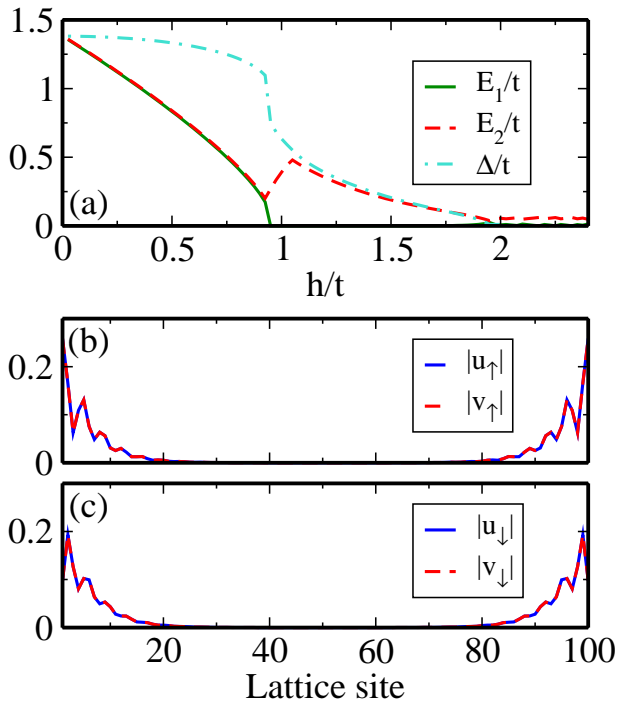}
\caption{(Color online) (a) Plot of the order parameter and the lowest two
eigenenergies $E_{1}$ and $E_{2}$ as a function of Zeeman field for the
transition from normal BCS superfluids to topological superfluids. $\protect%
\alpha =1.0,\protect\mu =-2.0$. (b,c) The Majorana zero energy state
wavefunction.}
\label{fig-edgestate}
\end{figure}

Different quantum phases in Fig. \ref{fig-noHS} can be characterized by
different density of states (DOS) $\rho(E)=\sum[|u_{i\sigma}|^2%
\delta(E-E_n)+|v_{i\sigma}|^2\delta(E+E_n)]$ and superfluid order parameter $%
\Delta \left( x\right) $, as shown in Fig. \ref{fig-dos}. In the insulator
phase (Fig. \ref{fig-dos}a), the order parameter $\Delta =0$ and there is an
energy gap for excitations. In the normal BCS superfluid phase (Fig. \ref%
{fig-dos}b), the order parameter is non-zero and there is a superfluid gap
around $E=0$. In the topological superfluid phase (Fig. \ref{fig-dos}c), a
zero energy peak appears in the DOS which corresponds to the zero energy
Majorana state. Note that for the normal BCS superfluid phase the order
parameter has a strong oscillation near each end, while for the topological
superfluid, the order parameters is a monotonic function of the sites.
Similar features have always been found for these two different phases for
different parameters. The FFLO phase (Fig. \ref{fig-dos}d) is characterized
by the spatially oscillating order parameter and the excitations are always
gapless.

\begin{figure}[tbp]
\centering
\includegraphics[width=3.2in]{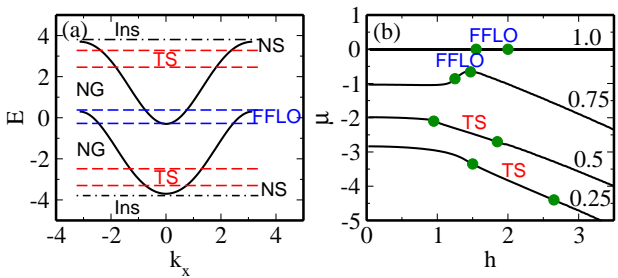}
\caption{(Color online) Single particle band structure of the spin-orbit
coupled optical lattices. $\protect\alpha =1.0$, $h=1.7$ (corresponds to the
dashed line in Fig. \protect\ref{fig-noHS}d). Corresponding chemical
potential regions for different phases are identified. (b) Plot of $\protect%
\mu $ as a function of Zeeman field for fixed filling factor $n=0.25$, $0.5$%
, $0.75$, $1.0$. The regions between the two circle symbols on each line
label the topological superfluid phase (for $n=0.25,0.5$) or the FFLO phase
(for $n=0.75, 1.0$).}
\label{fig-band}
\end{figure}

The emergence of Majorana zero energy state at the ends of the SO coupled
optical lattices can be clearly seen in Fig. \ref{fig-edgestate}. In Fig. %
\ref{fig-edgestate} we plot the bulk order parameter, the first and second
non-negative eigenvalues, denoted by $E_{1}$ and $E_{2}$, of the BdG
equation as a function of Zeeman field. We see that in the normal BCS
superfluid regime $E_{1}$ is always equal to $E_{2}$, and smaller than the
order parameter. The energy gap not equal to the order parameter is a unique
feature for spin-orbit coupled systems. When $h$ approach 1.0 ($h>\Delta $),
we observe a sudden jump of the order parameter and the system enters the
topological superfluid regime, where $E_{1}=0$, and $E_{2}$ increases and
reaches the maximum value 0.5 at $h\sim 1.1t$. When $h$ further increases, $%
E_{2}\sim \Delta $ gradually decreases and becomes zero when the system
enters the normal gas phase. In the topological superfluid phase, $E_{2}$ is
the minimum energy gap that protects the topological zero energy Majorana
state. In Fig. \ref{fig-edgestate}b and c, we plot zero energy state, which
is the eigenstate of the Bogoliubov quasiparticle operators $\Gamma
_{0}^{\dagger }=\sum_{i\sigma }(u_{i\sigma }^{0}c_{i\sigma }^{\dagger
}+v_{i\sigma }^{0}c_{i\sigma })$. The eigenstates satisfy that $u_{i\sigma
}=v_{i\sigma }$ on one end and $u_{i\sigma }=-v_{i\sigma }$ on the other
end. Thus by defining $\Gamma _{0}^{\dagger }=\gamma _{L}+i\gamma _{R}$, we
could identify the left and right end Majorana Fermions $\gamma _{L(R)}$.

Different phases in different parameter regions can be intuitively
understood from the single particle band structure, as shown in Fig. \ref%
{fig-band}a for $\alpha =1$, $h=1.7$ (corresponds to the dashed line in Fig. %
\ref{fig-noHS}d). The chemical potential regions for different phases are
identified in the figure. Comparing Fig. \ref{fig-noHS}d and Fig. \ref%
{fig-band}a, we see the topological superfluid phase appears when the
chemical potential cuts only a single band, while the FFLO phase appears
mainly around $\mu =0$ (filling factor $n=1$) and cuts two bands. The
insulator phase appears when two bands are either both fully occupied or
empty. Normal superfluid or normal gas phases also appear when a single band
is occupied with either small or large filling factors.

Because of the SO coupling, the spin polarization is not a conserved
quantity anymore, which is different from the spin-imbalanced Fermi gases in
the literature. However, the filling factor can still be controlled
precisely in experiments. With increasing Zeeman field, the chemical
potential changes for a fixed filling factor, leading to the transition
between different phases. In Fig. \ref{fig-band}b we plot the chemical
potential as a function of Zeeman field for different filling factor $n$.
Generally when $n<0.7$ or $n>1.3$, we find $\mu $ changes monotonically as a
function of the Zeeman field. However, in the case $n=0.75$, the chemical
potential does not change when $h<1$ and then increases slightly when the
system enters the FFLO phase regime. For $n=1$, the chemical potential is
independent of the filling factor due to the particle-hole symmetry. Notice
that $n$ is not a unique function of the chemical potential, therefore in
Fig. \ref{fig-noHS} we see that the FFLO phase is very large in the $h-n$
plane but becomes much smaller in the $h-\mu $ plane because the mapping is
a very complex function.

\begin{figure}[tbp]
\centering
\includegraphics[width=3.2in]{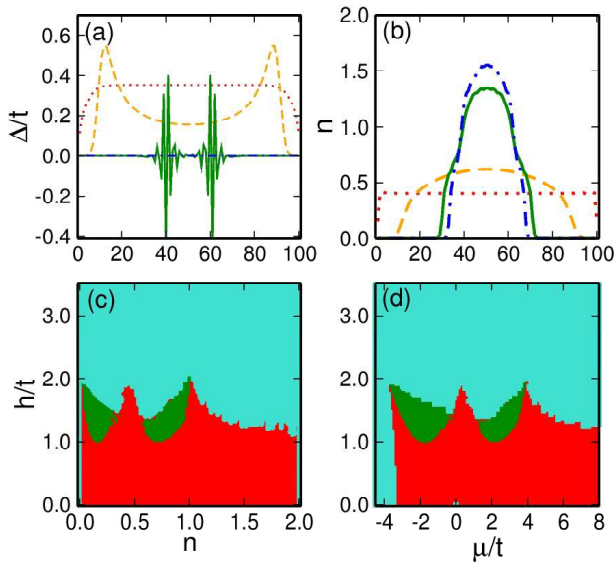}
\caption{Phase diagram in the presence of a harmonic trap. The order
parameter profiles (a) and atom density distributions (b) for different
trapping frequencies: $\protect\omega _{x}=0.0$ (Red dotted line), $0.05$
(Orange dashed line), $0.15$ (Green solid line), and $0.2$ (Blue dash-dotted
line). (c) and (d) are phase diagram as a function of Zeeman field $h$ and
filling factor $n$ or chemical potential $\protect\mu $. }
\label{fig-trap}
\end{figure}

\subsection{Phase diagram in a harmonic trap}

In a realistic experiment, a harmonic trapping potential $V(x)=\frac{1}{2}%
\omega _{x}^{2}(i-L_{c})^{2}$ exists, where $\omega _{x}$ is the trapping
frequency and $L_{c}$ is the center of the lattice. The effects of the
trapping potential on the filling factor and order parameter are shown in
Fig. \ref{fig-trap}(a) and (b). With increasing trapping frequency, the
ultracold atoms are forced to the center of the trap which has a lower
potential. With a high trapping frequency, a shell structure is developed,
where superfluid, insulator, and normal gas phases appear in different
regions of the harmonic trap. In the superfluid regime in the trap,
topological superfluid or FFLO phase may develop for different parameters.
In Fig. \ref{fig-trap} (c) and (d) we plot the phase diagram for a fixed
trapping potential. Notice that there is always a mixture of different
phases in the trap, therefore we only identify three different cases in our
plot: normal gas for the whole lattice, topological superfluids (with
zero-energy Majorana fermions) and other superfluids (normal BCS superfluids
or FFLO) in certain part of the lattice. In the topological superfluid, the
Majorana wavefunction does not localize at the two ends of the trap, but in
certain middle region of the trapping potential \cite{XJLiu2}. The position
of the Majorana fermion changes with the trapping frequency. With the
trapping potential, the effective chemical potential is site-dependent,
therefore the particle-hole symmetry is broken and the phase diagram is not
symmetric with respect to $n=1$ (Fig. \ref{fig-trap} (c)) or $\mu =0$ (Fig. %
\ref{fig-trap} (d)).

\begin{figure}[tbp]
\centering
\includegraphics[width=3.2in]{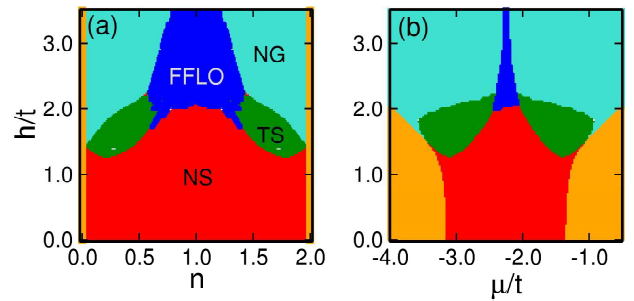}
\caption{Phase diagram with the inclusion of the Hartree shift. $\protect%
\alpha =0.5$. All other parameters and notations are the same as that in
Fig. \protect\ref{fig-noHS}.}
\label{fig-HS}
\end{figure}

\subsection{Effect of Hartree shift}

We also study the effect of the Hartree shift on the phase diagram. When the
Hartree shift is included, the local chemical potential and Zeeman field are
both modified as seen from $\tilde{\mu}_{i\sigma}=\mu+U\langle{n_{i\bar{%
\sigma}}}\rangle$. If we define $\langle n_i \rangle$ and $\langle m_i
\rangle$ as the local particle number and magnetization $m_{i}=n_{i\uparrow
}-n_{i\downarrow }$, then the modified chemical potential and Zeeman field
with including the Hartree shift term will be $\mu _{i}=\mu +U\langle
n_{i}\rangle $ and $h_{i}=h-{\frac{U}{2}}\langle m_{i}\rangle $. The
numerical results are presented in Fig. \ref{fig-HS} for a direct comparison
with the results without Hartree shift (Fig. \ref{fig-noHS} (c) and (d)). We
see the phase diagram is still qualitatively the same except that now the
phase diagram is symmetric for $\mu=-U/2$ as discussed in the above. Since
the Zeeman field $h_{i}$ depends strongly on the local magnetization $m_{i}$
and the FFLO phase is a competition between superfluid and magnetization, it
is hard to numerically determine FFLO phases near the phase boundary between
FFLO and topological superfluid phases. This is because the total free
energy becomes extremely complex as a function of order parameters and other
quantities, and most solutions we find correspond to excited states, instead
of global minimum of the free energy. Thus the boundary between topological
superfluid and FFLO phase cannot be determined precisely.

\subsection{Quantum phases in quasi-1D lattices}

\begin{figure}[tbp]
\centering
\includegraphics[width=3.2in]{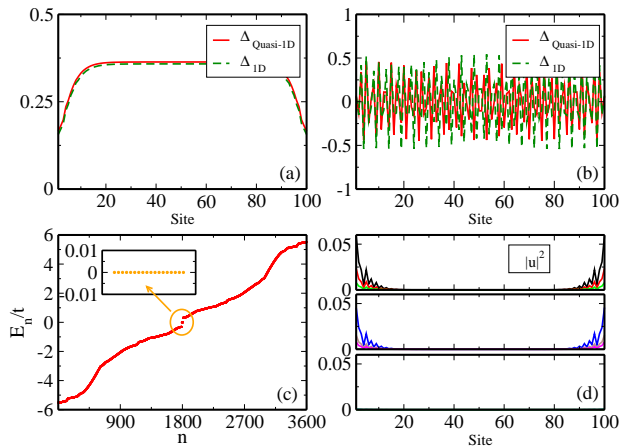}
\caption{Topological and FFLO phases in a quasi-1D ($100\times {3}\times {3}$%
) lattice. (a,b) The order parameters in topological superfluid (a) and FFLO
(b) phases. Red solid lines: quasi-1D; Green dashed lines: 1D for the same
parameters. (c) The BdG excitation spectrum in the topological superfluid
phase. The inset shows the nine zero energy states (18 shown due to the
finite size effect that lift the degeneracy slightly). (d) The corresponding
edge state wavefunction along nine tubes of the quasi-1D lattice for one of
the zero energy states. Parameters: $V_{z}=1.2$, $\protect\mu =-2$ for the
topological superfluid phase and $V_{z}=1.8$, $\protect\mu =0$ for the FFLO
phase.}
\label{fig-Q1}
\end{figure}

In a truly 1D system, quantum fluctuations become significant and need be
taken into account, which are beyond our mean field approximation. The
effects of quantum fluctuations can be suppressed by considering a
three-dimensional lattice with weak coupling $t_{\perp }=0.1t$ along two
transversal directions, similar as the high temperature cuprate
superconductor where 2D superconductivity is stabilized by weak coupling
along the third direction. In experiments, such setup can be realized in 3D
degenerate Fermi gases subject to 2D strong optical lattices, forming weakly
coupled 1D tube arrays. The 1D weak lattice and spin-orbit coupling are then
applied on each tube. The quantum phases in the quasi-1D system are
calculated by self-consistently solving the corresponding BdG equations and
are found to be similar to the above 1D results. Here we focus on the
interesting topological and FFLO phases. The order parameters of these two
phases in quasi-1D lattices are shown in Figs. \ref{fig-Q1} (a) and (b).
Comparing to 1D results, the order parameters only change slightly due to
the week tunneling between tubes. In contrast to 1D lattices, the array of
lattice tubes can host multiple Majorana zero energy states at the ends of
each tube, as shown in Fig. \ref{fig-Q1} (c) for the Bogoliubov excitation
spectrum. However, the weak tunnelings do not lift the zero energy
degeneracy because they are along the traversal directions, which are
different from the strong energy splitting induced by the coupling between
two Majorana fermions in the same tube due to the short tube length. The
zero energy wavefunctions are still localized around the edges of the tubes,
but can expand along the transversal directions due to the weak tunneling,
as shown in Fig. \ref{fig-Q1}d.

\label{sec: Quasi1D}

\section{Conclusion}

\label{sec: summary}

In summary, in this paper, we address the question that which phases, FFLO
or Majorana superfluids, will survive in spin-orbit coupled Fermi gases in
the presence of a large Zeeman field through studying the mean field quantum
phases of 1D and quasi-1D spin-orbit coupled optical lattices. In the
optical lattice system each site can host at most 2 fermions, making the
system host plenty of phases depending on the filling factor and the Zeeman
field, which are quite different from the free space results. At a finite
Zeeman field we observe the strong competition between topological
superfluid phase and FFLO phase. The SO coupling enhances the topological
superfluid phase while suppresses the FFLO phase. The weak tunneling along
transversal directions in quasi-1D lattices does not change the results.
These results are not only important for the search of Majorana fermions in
spin-orbit coupled degenerate Fermi gases, but may also have significant
applications for the solid state nanowire heterostructures where similar
physics exists.

\textbf{Acknowledgement:} The authors acknowledge the supported by ARO
(W911NF-12-1-0334), AFOSR (FA9550-13-1-0045), and NSF-PHY (1104546).

\end{document}